\documentclass[amsmath,amssymb,aps,twocolumn]{revtex4}

\usepackage{amsmath}
\usepackage{amssymb}
\usepackage{amsthm}
\usepackage{psfrag}
\usepackage{graphicx}

\newcommand{\be}{\begin{equation}}
\newcommand{\ee}{\end{equation}}
\newcommand{\bea}{\begin{eqnarray}}
\newcommand{\eea}{\end{eqnarray}}

\input epsf%
\begin{document}
\title{On Asymptotic Darkness in Ho\v{r}ava-Lifshitz Gravity}
\author{Emilio Elizalde$^a$ and Pedro J. Silva$^{a,b}$}
 \affiliation{$^a$ICE/CSIC and IEEC, Universitat Aut\`onoma de Barcelona, 08193 Bellaterra, Barcelona Spain\\
 $^b$Departament de F\'isica and IFAE, Universitat Aut\`onoma de Barcelona,
 Bellaterra 08193 Barcelona Spain}

\begin{abstract}
Ho\v{r}ava-Lifshitz gravity is shown to exhibit a peculiar behavior, when scattering amplitudes are considered. At low energies it seems to classicalize i.e. the effective size of the interaction grows as a function of the $s$-parameter, with BHs forming part of the spectrum; but when the probing energy is increased such that higher order operators become important, this behavior changes and the classicalon recedes to a new configuration where ordinary quantum regimes take over. Eventually, the theory behaves as a usual field theory that allows probing arbitrarily small distances. In particular, the classical potential created by a point-like source is finite everywhere exhibiting a Vainshtein alike screening behavior. The transition from one behavior to the other is carefully described in a particular case.
\end{abstract}


\maketitle

\maketitle

\noindent {\it 1.~Introduction.}---General Relativity (GR) is not perturbatively renormalizable. If there is at all a local quantum theory of gravity (QTG), based on covariance and Lorentz invariance (LI), which reduces to GR in the infrared (IR) regime, it cannot be a standard field theory.\footnote{Namely, a local field theory that is LI and follows the principles of quantum mechanics. These theories, are well defined provided that, in the ultraviolet regime, they are Gaussian theories admitting, at most, perturbations by relevant operators.} This is based on the universal character of gravitational collapse in GR, guaranteeing an ultaviolet (UV) spectrum dominated by black hole (BH) states, often refereed to as {\it asymptotic darkness}. This means, in turn, that the entropy of our sought for QTG should necessarily grow as the area, and not as the volume of the spacetime region (as in normal field theories).

Present candidates for a QTG  are string theory, loop quantum gravity, the so-called asymptotic freedom approach and, more recently, the Ho\v{r}ava-Lifshitz theory (HL). Only the first and the last proposals in this list might cope with the syllogism of the preceding paragraph, since string theory is in fact a non-standard field theory, while HL explicit breaks Lorentz invariance. We want to learn what this last proposal can tell us about the gravitational degrees of freedom which propagate in the deep UV regime and its relation to a putative BH in the theory. In other words,
we are asking the question
if there is asymptotic darkness in HL.

HL seems renormalizable, at least at the level of power counting. This is seen by introducing irrelevant operators which explicitly break LI but ameliorate the UV divergences. Also, Lorentz invariance is expected to be recovered at the deep IR regime as an accidental symmetry, but has no intrinsic axiomatic character.
In fact, this mechanism is by no means strange to condensed mater field theory, which counts with many examples thereof (actually realized in nature), where a system is described in the UV by a non relativistic field theory that nevertheless runs into a relativistic field theory in the IR.
The above phenomena were first studied by Lifshitz in connection to solid state physics. The original proposal of \cite{H} has been enriched in different ways. Here, we deal with the modification of \cite{Sotiriou:2009bx-gy,Blas}, a most promising version of the theory which seems to be able to accommodate both theoretical and experimental tests.
\medskip

\noindent {\it 2.~The HL model.}---The HL gravitational dynamical variables are taken to be the laps $N$, shift $N_i$, and space metric
$g_{\,ij}$, the Latin index running from 1 to 3, characteristic of the ADM construction:
%
$ ds^2=-N^2dt^2+g_{ij}(dx^i+N^i)(dx^j+N^j), \, $
%
with $N^i=g^{ij}N_j$. The action is written in terms of geometric objects, normally defined in the ADM slicing of space-time, like the 3d-covariant derivative $\nabla_i$, the spatial curvature tensor $R_{ijkl}$, and the extrinsic curvature $K_{ij}$. The HL action is
\bea
\int dt dx^3\,N\sqrt{g}\left({\cal L}_{kinetic}-{\cal L}_{potential}+{\cal L}_{matter}\right),\;
\eea
where the kinetic term is universally given by
%
$\label{k} {\cal L}_{kinetic}=\alpha(K_{ij}K^{ij}-\lambda K^2),\, $
$\alpha$ and $\lambda$ playing the role of coupling constants. The potential is a generic function of $(R_{ijkl}$, $\nabla_i$, $\nabla_i \ln(N)\equiv a_i)$.
The action generically breaks covariance down to the subgroup of 3-dimensional diffeomorphism invariance and time reparametrizations: $x\rightarrow \tilde x(t,x)$, $t\rightarrow \tilde t(t)$. Assigning dimensions $-1$ to space and $-3$ to time, it can be proven that it is enough to restrict the potential to be constructed out of operators up to dimension 6, in order to get a power counting renormalizable theory.

In general, the potential can be written implicitly as
\be {\cal L}_{potential}=\alpha \sum
\beta_n\,O^n(a_i,\nabla_j,R_{ijkl}),
\label{bps}
\ee
where $O^n$ stands for a general operator of maximum dimension 6 and $\beta_n$ is the corresponding coupling constant \cite{Carloni}.
 Here, we just write down some of the possible terms, to illustrate the general form of the potential,
\be
\beta_1R+ \beta_2 a^ia_i+\beta_3 R^2+\beta_4a_i\Delta a^i+\beta_5\Delta R \nabla_ia+ \ldots \, ,
\label{pot}
\ee
where $\Delta=\nabla_i\nabla^i$. The matter Lagrangian should be written in terms of general couplings between matter fields and the metric variables. In contrast to GR, there is now no argument based on symmetries, to choose the minimal coupling over more exotic options. Nevertheless, since phenomenological observations do seem to favor it, we will here use this option in our discussion.\footnote{Matter couplings in HL represent an unsolved problem; see, for example, \cite{Carloni} for  possible approaches.}

At low energies
this model reduces to a particular family of Einstein-aether theories, with the aether chosen to define a hyper-surface orthogonal space-time foliation, which we call EAO \cite{Jacobson:2010mx}. There is here an extra mode, a propagating scalar, on top of the familiar helicity-2 graviton.
Therefore, we can use our knowledge of these theories---extensively studied, with many observational constraints having been incorporated to the model \cite{Jacobson:2008aj}---to better understand the structure and phenomenological implications of them on HL theories. Nevertheless, keep in mind that this is purely an IR relation, hence there may be important differences when extrapolating results to the UV regime.\footnote{For example, the velocity of propagation of the scalar mode in the HL case is unbounded from above, in the UV regime, while in the EA case, this velocity is superluminal, but finite.}
\medskip

\noindent {\it 3.~HL black holes in the IR regime.}---Here we study low energy configurations in HL that are found in EAO. Even though EAO breaks LI and there are different superluminal propagating modes, apart from the helicity-2 graviton, it is possible to find BH solutions that have an interior region, causally isolated from the exterior. 
In these solutions the notion of ``event horizon" has to be generalized, to capture the phenomenology of all different propagating modes appearing in the pure gravitational sector. In fact, each different mode censors a different ``effective metric" and, therefore, has as associated a different event-horizon. In \cite{Barausse:2011pu}, a one parameter family of spherically symmetric, static, and asymptotically flat solution was found. These BHs have a regular metric horizon and a scalar field horizon such that, in general, the ``metric horizon" is located outside of the ``scalar field horizon". 
The exteriors of these BHs are very similar to Schwarzschild BHs in GR, where a single parameter, identified with the total mass, characterizes the solution. Interiors, on the contrary, do change more dramatically, since certain characteristic functions constructed from the metric oscillate as they approach the space like singularity in the center, always producing a point of no-return back to space-like infinity. This happens regardless of the nature and the velocity of the propagating mode. The point of no return is located, in general, very close to the interior of the scalar horizon defining a ``universal horizon".

It is not yet proven that such BHs are the final stage of gravitational collapse.
This is indeed a hallmark in GR, crucial for the reasoning behind AD, but it is yet unknown whether in EAO  such is generically the case. For some initial conditions it was found in \cite{Barausse:2011pu}  that  the above regular BHs do form as final stages of gravitational collapse and are stable under general perturbations. 
Another important feature of a BH is its associated thermodynamics. In GR, a BH is endowed with energy, temperature and entropy, proportional to its area. These observables obey zero, first and second laws of thermodynamics, defining altogether the so-called BH thermodynamics.
Further, all these concepts have been generalized to a local off-shell version of ``thermodynamics of space-time", where Einstein's equations are recovered as an equation of state \cite{Jacobson:1995ab} of equilibrium thermodynamics. This remarkable result has been demonstrated to apply not only to GR, but also to a large family of generalizations of gravity, known as f(R)-gravities \cite{Elizalde}. Space-time thermodynamics  has its most clear microscopic derivation in string theory, where for some supersymmetric  BHs it has been shown that the entropy  is equal to the logarithm of the degeneracy of the configuration that collapsed into the BH in the first place, what anticipates the possibility of constructing a rational foundation of GR (see \cite{Silva} and articles therein).
In EAO the situation is different. As already said, due to the breakdown of LI one generically has different fields propagating at different speeds and, as the different fields have different horizons they actually feel different Hawking temperatures. These BHs are somehow out of equilibrium and could be used, in principle, as perpetual motion machines of the second kind. In other words, the second law of thermodynamics is most probably violated here \cite{Jacobson:2008yc}. For similar reasons, it is not yet known how to define the BH entropy itself. Thus, the whole idea of thermodynamics in an EAO BH is far from being clear, at present.

Based on the above,  it does seem that in HL
BHs could form, specially after realizing that we have a sort of universal horizon, which would clearly define an isolated causal region (since in HL the velocity of propagation grows unboundedly with the energy; see e.g.~\cite{Blas}). Unfortunately, the above conclusion is incomplete since we have not included the higher-order operators appearing in HL, which could modify the behavior of the interior of the solution. Notice that there are solutions in HL which are not solutions in EAO, although the contrary is not true. These extra operators, not appearing in EAO, influence the propagation of scalar and matter sectors. This issue has been partially studied in \cite{poly}---using the appropriate generalization of geodesic equations---with the result that the notion of Hawking temperature is particle-dependent, yielding a non universal definition of temperature. If particles have dispersive geodesics, the will-be Hawking radiation ceases to be thermal \cite{corley}. In fact there appears to be a strong correlation between LI and the existence of a BH thermal radiation (see \cite{Jacobson:2008yc}). Ultimately, the thermality of the Hawking radiation is related to the Unruh effect, which heavily relies on LI.

Therefore, although it does seem in principle possible to have a notion of BHs in HL, it is also clear from the preceding analysis that these objects would be rather different from GR BHs (as there might be no thermodynamics associated to them and it is not clear how universal its formation is). At this point we need another method to better asses the importance of the higher dimensional operators in HL, which could be important for the dynamical formation of BH, as we approach the UV and strong curvature regimes.
\medskip

\noindent {\it 4.~UV degrees of freedom in HL.}---BH formation is an involved non-linear process which, most probably, can only be tackled down using numerical methods. Even then, owing to the so many different operators appearing in the theory, its complexity is enormous. We need to consider complementary approaches if we want to understand the process. There is a method most adapted to a particle physicist's mind, where the scattering of the relevant propagating degrees of freedom is studied, in the deep UV regimes, at small enough impact parameter (in this way one brings a lot of energy within a very small volume, what is a natural form to probe small distances to uncover the UV degrees of freedom).
Calculating scattering amplitudes is easy, since one relies on linear perturbation theory on a given vacuum, where the asymptotic states are free. In GR, once the  concentration of energy lies inside its Schwarzschild radius, one always expects the formation of a BH. BH formed in a scattering experiment can be recognized, in the linearized approach, by the breakdown of the perturbation expansion in the gravitational field. Such field configuration is viewed as a precursor of the full, non-linear BH.

This idea has recently been used to study the UV self-completion of GR (and other field theories of Goldstone-scalar type) in a non-Wilsonian sense, termed {\it classicalization} \cite{Dvali1}.  Classicalization means that a seemingly unitary-violating theory prevents from going to sub-cutoff distances, by becoming classical in the deep UV. A characteristic of it is the existence of energy self-sourcing and the emergence of a classical scale ${r_*}$ which prevails over any other quantum scale at sufficiently high energies (one finds an UV/large-distance relation).


The proper definition of $r_*$ is naturally given as the distance below which the scattering process at a given center of mass energy $e$ cannot be ignored. The scattered field $\phi$ can be written as a superposition of the free one $\phi_0$ and a scattered part $\phi_1$, where $\phi_1<<\phi_0$ asymptotically towards past infinity, but it increases up to a point where $\phi_1 \sim \phi_0$, as time evolves. At this point the perturbative expansion, based on small perturbations of $\phi_0$, breaks down defining the region where the gravitational interaction becomes dominant. The key idea behind classicalization is the fact that $r_*$  will eventually become the dominant scale, over any quantum scale, if it grows with $e$. In this case, the result of the quantum scattering is an extended classical configuration, called {\it classicalon}, which hides microscopic distances away. 


The main conclusion, when applying these ideas to GR, is that the scattering of gravitons at energy $e$ produces indeed a classicalon with $r_*$ equal to the Schwarzschild radius $\l_p^2 e$ (where $l_p$ is Planck's length). This is observed for two-graviton scattering and for one graviton in the presence of an external source. This configuration is surely a precursor of the full BH, based on Birkoff's theorems. Also, at fixed $e$, GR has the largest $r_*$ when compared to other LI classicalizing theories, like Goldstone-scalar ones and, in particular, to any covariant LI modification of gravity that reduces to GR in the IR \cite{Dvali1}.  In other words, it looks as if GR were self-complete in the UV since, as stated by AD, the UV is populated by BH states which take care of the unitarization of the theory. 

We will now use the above method to study the UV states in HL and its possible relation to putative BH formation. First, we consider the case of EAO (that corresponds to the IR regime), to then apply this same machinery to a completed action with higher-order terms, to capture the UV behavior of the HL theory. We aim at understanding better the nature of the gravitational collapse in both limiting theories, EAO and HL, identifying thereby the main key differences due to the inclusion of higher-order operators.
Although in HL there are spin-2 and scalar propagations, we will only concentrate in the scalar response to an external current, since this is the simplest calculation that, nevertheless, contains all the main ingredients and hints to results of other possibilities involving more complicated scattering with helicity-2 gravitons (what will be done in a forthcoming more detailed article).
Here, we first present a calculation for the classical potential generated by a static source, to then consider the fully fledged s-wave self interacting scattering \footnote{Notice that the mass, $m$, is to be related to the s-parameter in the laboratory frame only for low energy scattering.}.

The scalar perturbation of the flat space-time metric, adapted to static configurations (written always through use of the HL parametrization) is given by the expansion
\be
N=1+\phi,\quad g_{ij}=\delta_{ij}-2(\delta_{ij}-{\partial_i\partial_j\over \Delta})\psi\,.
\ee
For EAO, the relevant part of the Lagrangian coupled to the simplest external source, a point-like object of mass $m$, is
\bea
{\cal L} = \frac{1}{2l_p}\left(-2\psi\Delta \psi +4\phi\Delta\psi -\alpha\phi\Delta\phi\right) -\phi m\delta^3({\bf x})\,.
\eea
Solving the field equations, we  get that $\psi=\phi$,
\be
\Delta \phi=\frac{ml_p^2}{(2-\alpha)}\delta^3({\bf x})\quad \Longrightarrow\quad \phi = -{ml_p^2\over 8\pi(1-\alpha)r},
\ee
where, in this gauge, $\phi$ corresponds to the Newtonian potential. This result describes a gravitational field growing monotonically as we approach the external source. Notice that the gravitational field sums up to an infinite wall potential, what would eventually translate into a breakdown of the linearized approach. A key observation is that, for any test observer  probing the system, the value at which the linearized approach breaks down defines  the ``size of the interaction", $r_*$, which turns out to be proportional to the mass of the source, $m$. Owing to the monotonicity of the potential, this is true for any energy range we use to probed the source. Therefore, we get
\be
r_* \sim \frac{ml_p^2}{(1-\alpha)}.
\ee
As we increase $m$, the size of $r_*$ grows up and will eventually dominate over any other quantum scale in the theory. This is the hallmark of classicalization, which tells us that in EAO there is a minimal length that we can probe by a scattering process. Indeed, any attempt to go further will instead produce larger and larger classical configurations which will take us back to the IR domain of the theory.
Notice that, owing to the symmetries in this ansatz, the resulting $r_*$  is similar to the  Schwarzschild radius $l_p^2 m$.
This result is consistent with the IR dynamics, since what we have found should be understood as the precursor of the BH solutions in EAO. For one, the fact that the scalar horizon is always within the helicity-2 horizon can be understood as a consequence of classicalization of the helicity-2 state being more efficient than for any other propagating field, in particular, the scalar field.


Our next step is to include in the above calculation higher-order operators, what marks the departure of HL from EAO. With the same ansatz, the new terms added to the EAO Lagrangian are
\bea
  \frac{1}{2l_p}\left[  -f_1(\Delta\psi)^2-2f_2\Delta\phi\Delta\psi -f_3(\Delta \phi)^2\right. \nonumber \\
\left. -g_1\psi\Delta^3 \psi -2g_2\phi\Delta^3\psi -g_3\phi\Delta^3\phi \right]\,.
\eea
The associated field equations are more complicated since higher derivative operators are present. For our purposes, it is better to work in momentum space, $k_i$, where the equations are  easier to solve for both fields, $\psi, \phi$, yielding the result in terms of three polynomials in $k^2$:
\bea
&\psi(k)=\frac{A(k)}{B(k)}\phi(k)\,, \quad \phi(k)=-\frac{m \, l_p^2 B(k)}{2[A(k)^2+B(k)C(k)]k^2}&\nonumber\\
&A(k)=1-f_2k^2-g_2k^4\,,  \quad
B(k)=1+f_1k^2+g_1k^4\,,&\nonumber\\
&C(k)=1+f_3k^2+g_3k^4\,.&
\eea
%
Here we use the same symbol for the fields in momentum space.
Observe that in the small $k^2$ cases, where all the the polynomial functions $A,B,C$ reduce to the identity, we recover our previous  result for EAO. However, in the other extreme regime, where the polynomials are dominated by higher powers of $k^2$, the result is very different: $\phi$  grows, generically, with a power six of the momentum, that corresponds to a cubic power of the space distance $r$.  In other words, the resulting potential is not any more a monotonic function of $r$, in fact is not difficult to see that in the general case the potential {\it does not diverge} at the source! This is a beautiful result since it implies that the gravitational response gets regularized at short distances. What we obtain is a sort of screened potential (Vainshtein screening) that has a very different response at large distances from the one at short distances.
The similarity of the situation here with the one encountered in Galileon models \cite{galil1} is very remarkable.
The analysis of the associated scattering process in this case turns out to be more involved since for low energy regimes one will feel the potential related to the large distance response while, for higher energy regimes, one is able to overpass the initial potential wall, of the form $1/r$, to access the short range form of the potential, i.e. a cubic power of the space distance $r$.

\begin{figure}[htb]
\begin{center}
\includegraphics[width=80mm]{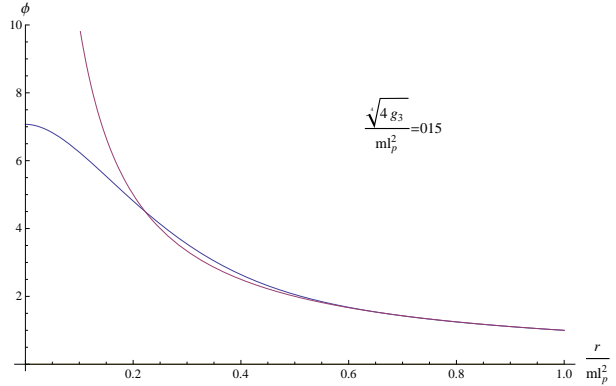}
\end{center}
\vskip-7mm
\caption{ {\protect\small
Classical potential for an external source as a
function of $r$. The purple line corresponds to the EAO case with usual
$1/r$ dependence, while the blue line corresponds to the potential
explicitly calculated in the text, showing the characteristic screened
potential.}}
\label{f1}
\end{figure}

The long distance behavior coincides with the EAO case, already discussed before, so let us study the short distance behavior. In this case, once the probe has enough energy to overpass the initial potential wall related to the $1/r$ asymptotic behavior, the potentials is proportional to $r^3$, so that $r_*\propto 1/\sqrt[3] {m}$, giving no increase of $r_*$ with the energy of the source! In this scenario the above approach shows, therefore, {\it no classicalization} at all. To see this more clearly, consider a particularly simple case that exhibits the generic behavior, where among the different couplings only $g_3$ be non-zero, so that our solution simplifies to
\bea
\psi=\phi\,,\quad \phi = -\frac{ml_p^2}{2(2k^2+g_3 k^6)}.
\eea
 In space variables the above expression yields
 \bea
\phi(r)\sim -{m l_p^2 \over r} \left[1 - e^{-r/ \sqrt[4]{4g_3}} \cos(r/\sqrt[4]{4g_3})\right]\,.
\eea
Hence, for little energy probes, only the large distance behavior is felt, corresponding to the $1/r$ tail of the potential, while for high energy probes, the short distance behavior is effectively felt, corresponding to the $r^3$ dependence of the potential. Again, {\it no classicalization occurs}, since $r_*\propto 1/\sqrt[3] {m}$, which diminishes when we increase the energy of the external source (see Fig.~\ref{f1}).

To check the above conclusion further, let us compute $r_*$ from breakdown of the linear approximation on a scattering of the self-interacting scalar fields, $\phi$, of the full HL action. The relevant part of the Lagrangian is a simple extension of the one considered before, modified in order to include the necessary time-dependent part, namely
\bea
{\cal L} = \frac{1}{2l_p}\left[-2{(3\lambda-1)\over(\lambda-1)}{(\partial_t\phi)}^2+\phi {\cal M} \phi \right]  \,,
\eea
where  ${\cal M}$ is a polynomial operator that depends on $\alpha,f_i,g_i$ and $\Delta$. For simplicity we will consider the case where only $f_3$ is different from zero, that will be enough to illustrate the generic behavior. In this case, $M={2\over\alpha}[(2-\alpha) - (3-\alpha)f_3\Delta]\Delta$ and the field equations can be recast as
\be
\Box \phi -F_3\Delta^2 \phi =0.
\ee
Here we have rescaled both time and space to recover the Laplacian and $F_3=\alpha{(3-\alpha)\over (2-\alpha)^2}f_3$ \footnote{Recall that, from the analysis of \cite{Blas}, the parameters $\alpha,\lambda$ are restricted to the ranges $0<\alpha<2$ and $3\lambda-1>\lambda-1$}.

We now iteratively solve the above self-interacting equation in perturbation theory, by writing $\phi=\phi_0+\phi_1$, where $\phi_1$ is a small perturbation of $\phi_0$. Then, at zeroth and first order, we get the equations
\be
\Box \phi_0 =0 \quad \hbox{and} \quad \Box\phi_1=F_3\Delta^2\phi_0\,.
\ee
The scattering process we are considering is such that, at $r=\infty$ and $t=-\infty$, our scalar field is well approximated by a spherically symmetric plane wave $A e^{w(t+r)}/r$, solving the zeroth order equation,  of high center of mass energy $\sqrt{s}\sim w$, while the corresponding solution to the first order equation is of the form $AF_3w^3(r-t) e^{w(t+r)}/r$. As explained previously, $r_*$ corresponds to the region where the perturbation approach breaks down, i.e. where $\phi_0\sim \phi_1$. In our case, we get
\be
r_*\sim {1\over F_3w^3} \Longrightarrow r_* \sim {1\over (\sqrt{s})^3 }\,.
\ee
Therefore, the scattering calculation shows that $r_*$ decreases as we increase the energy of scattering, that reveals no signal of classicalization.

The above results mean that HL exhibits a peculiar behavior when scattering amplitudes are considered. Indeed, at low energies, it seems to classicalize with effective BHs forming part of the spectrum, but as we increase the probing energy, the illusion just vanishes, and the classicalon recedes to a new situation where ordinary quantum regimes take over. In other words, the theory behaves as a usual field theory in the deep UV, that allows probing arbitrary small distances. This mixed behavior, where we start with asymptotically free quantum states, which we scatter to form, at low energies, classical configurations, but which, at high energies, lead back to quantum states, is known as  {\it de-classicalization} \cite{Dvali3}. De-classicalization is not at all possible in GR and its covariant LI generalizations due to the insurmountable strong constraints imposed by the positivity of the energy and non-existence of ghost states. We have proven that HL provides a completely different scenario where the aether breaks LI.

A consequence  of de-classicalization in HL is the non-existence of any precursor of a BH to be formed in the deep UV regime. This makes it difficult to believe in the eventual existence of BHs at all. The only valid conclusion  in this case appears to be that BHs are just an {\it illusion at low energy}, which is removed once we include into the system higher momentum states.
\medskip

\noindent {\it 5.~Discussion.}---High energy scattering in HL leads to de-classicalization. Therefore, there is no precursor to BH formation in this scenario, which leads to a behavior characteristic of renormalizable field theories where, in principle, one can probe any arbitrary small distance. In this regard, we see no UV/IR relation. The theory is therefore very different from GR, where the UV spectrum is characterized by classical BHs, which implement an insurmountable barrier to probe distances below the Planck length, yielding the well-known UV/IR characteristic relation.  As a consequence, already at this level we get no asymptotic darkness in HL.

On top of the above, a natural consequence of de-classicalization is that BH formation should not take place, for it is difficult to understand BH formation without the existence of an accompanying  precursor. Even in the low energy regime, where the EAO description is correct, there seems to be no well defined area law for the entropy of BH configurations and, therefore, we are left with no argument to question the entropy behavior of the HL theory in the deep UV regime, what is a key point of the asymptotic darkness syllogism.

Notwithstanding that, as a byproduct of our analysis we have found a pleasant surprise, in that EAO theories do classicalize, yielding a well defined precursor to BH formation. This information, complemented with the encounter of BHs in the full non-linear regime \cite{Barausse:2011pu}, produces a self consisting picture where, even though LI is broken, there persists a notion of BH with a universally causal trapped-surface. In this theory there is indeed a self completion mechanism at work (as there is in GR) with an UV/large distance relation which serves to unitarize scattering amplitudes in a non Wilsonian manner. In other words, EAO is as good an option as GR is, at least at the theoretical level.

Finally, de-classicalization in HL seems to be in tension with the existence of BHs in EAO, since those exhibit a universal horizon, regardless of the energy of the probe we are considering. A possible way out is that these solutions may change dramatically once the full HL theory is considered. Recall that such BHs where found by imposing the existence of a horizon, in the first place, to then complete the metric towards the center of the solution. The result is an oscillatory behavior and a space-like singularity. But a complete theory of space-time should not have singular points. This singularity could be telling us that the ansatz is non-physical and that, once the singularity has been taken care of, in the full HL theory the oscillatory behavior could just disappear. In any case, this issue demands further investigation in order to be fully understood.

\vspace*{3mm}

\noindent{\bf Acknowledgments.} We thank G. Dvali and O. Pujol\`as for useful discussions.
This work has been supported by UAB,  EU ITN (PITN-GA-2009-237920), MIUR (2006022501), MICINN (Spain), projects FIS2006-02842 and FIS2010-15640, the CPAN Consolider Project, and AGAUR (Generalitat de Ca\-ta\-lu\-nya), contract 2009SGR-994.
EE's research was performed in part at Dartmouth College, NH, USA.


\end{document}